# An Efficient Technique for Text Compression

Md. Abul Kalam Azad, Rezwana Sharmeen, Shabbir Ahmad and S. M. Kamruzzaman [1]
Department of Computer Science & Engineering,
International Islamic University Chittagong,
Chittagong, Bangladesh.
[1] Department of Computer Science & Engineering,
Manarat International University,
Dhaka, Bangladesh.
Email: {azadarif, r_sharmin_79, bappi_51, smk_iiuc}@yahoo.com

**Abstract**

For storing a word or the whole text segment, we need a huge storage space. Typically a character requires 1 Byte for storing it in memory. Compression of the memory is very important for data management. In case of memory requirement compression for text data, loseless memory compression is needed. We are suggesting a lossless memory requirement compression method for text data compression. The proposed compression method will compress the text segment or the text file based on two level approaches firstly reduction and secondly compression. Reduction will be done using a word lookup table not using traditional indexing system, then compression will be done using currently available compression methods. The word lookup table will be a part of the operating system and the reduction will be done by the operating system. According to this method each word will be replaced by an address value. This method can quite effectively reduce the size of persistent memory required for text data. At the end of the first level compression with the use of word lookup table, a binary file containing the addresses will be generated. Since the proposed method does not use any compression algorithm in the first level so this file can be compressed using the popular compression algorithms and finally will provide a great deal of data compression on purely English text data.

## 1. Introduction

A text segment is a collection of words and a word consists of characters. All the characters are unique and they are the basic units of word. That's why to store a text segment it is needed to store all the words separately. To store a word, all the characters that the word contains needed to be stored. For this type of storing mechanism a huge amount of disk space is needed and in the current world this technique is used. But this type of system of storing text makes the text segment consume more space. Let us suppose that a text segment contains some word "n" times, which is 7 characters in lengths (take it as average length) then for repeated presence of the same word of "n" times it need n*7 Bytes. If some sort of indexing within the text segment is done, then the text segment size can be reduced. But this process is still not effective because it needs extra space to make the indexing table and sometimes it may increase the file size rather than decreasing. Now if a word index table which will be used to index the text segment for reducing text segment size can be made and the index table will not be a part of the text segment but rather will be a part of the operating system then it will be an effective way to reduce the text segment memory requirement size. This proposed method decreases the persistent memory requirement approximately 50% and more. In this process it will generate a binary file.

After the reduction work, the traditional compression will be done over the binary file. The compression characteristics of the file is, as the file generated the machine through lookup table address has already decreased the memory requirement of the text data and generated a binary file then if Deflate algorithm is carried over the file, which is the improved form of Huffman coding algorithm and LZ-77 coding algorithm, can reduce the memory requirement quite dramatically.

The lookup table is a special related table in which numeric item values are classified into categories. An INFO lookup table contains at least two items: the relate item and an item named either symbol or level. In our proposal the lookup table will have an address value of a word. In the time of memory requirement compression the word in the memory will be replaced by that value. Since the value needs a small amount of bits to represent itself, thus it will reduce the storage requirement.

As any language is not a rigid body that is. Languages are always expanding and day by day is enriching with the invention of new words and also by adoption of new words. So some challenges were faced in choosing the boundary of languages. More over the following things had to be considered.

## 1.1. Spelling mistake

The base of the our proposed methodology is the words in English dictionary but due to spelling mistake there may be lots of mistakes which will not be matched with no entries in the dictionary.

## 1.2. Technical and biological term

Technical word, Chemical names and names of different species of animals is in amount more than 1 millions but those words are not generally added in the dictionaries. So it is needed to decide whether those names will be allowed in the lookup table.

## 1.3. Other language words

Sometimes words from other language that have not been yet adopted in the English may be found in the text segment. Though the word may be common in use it will be considered as a special word that is a constant word.

## 2. Proposed method

The proposed compression will be carried of the basis of two leveled approach. In the first level the text will be reduced using word lookup table then in the second level text will be compressed using Deflate algorithm.

## 2.1. Text reduction

To reduce the size of the text segment the proposed approach will use a word lookup table, which will be acting like a word store and each particular word will be assigned an address value and any particular word will be determined by that value. For that reason, during the time of storing the word, the word will not be stored as a Byte stream based on each of the character's ASCII value, rather it will be stored as a fixed bits size long bits stream that will form the address value which will reference the address of the text in the word table.

**2.1.1. Word lookup table.** A word lookup table is a special tabular data file containing the text dimension of a word as an attribute of an address, which is used to pop up text to display the possible text data for a field. For our purpose we have decided to use a 19-bit word lookup table. A 19-bit word lookup table can contain a number of,

2^19 = 524,288 entries.

That is if a 19-bit lookup table is used then it can index a number of 0.524 millions of different words. Where is in current English language there is an approximation of having a total of 0.470 million of words [3, 6]. It shows that by using a 19-bit lookup table it can easily index any English text. As there is only 0.470 millions of different word in English language that is there are still about (0.524-0.470) = 0.054 Millions of empty entries. That is after deducting some entries for special situation handling there are approximately 52,000 entries empty. These empty entries will be used for farther improvement on the proposed methodology. The block diagram of the word lookup table is shown below in Figure 1:

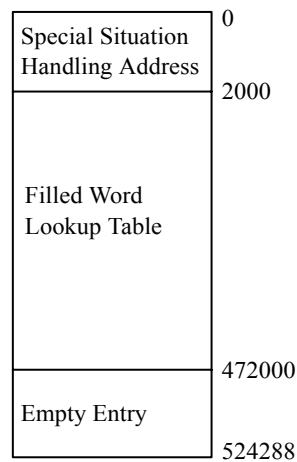

**Figure 1:** Block diagram of the word lookup table

**2.1.2. Word storing architecture.** From experiment it was found that English language word consists of 6.91 characters in an average. That's why, in the word lookup table, the size of any word is taken 7 characters long in an average. The storing architecture of the word will be shown below in Figure 2:

| Index Address | Word Stored in Details | Ending signal |
|---|---|---|
| 19-bit Address | 7 Characters Long Word | Ending signal |
| 19 bit | 6 * 7 = 42 bit | 6 bit |

| Total 67 bit |
|---|

**Figure 2:** Word storing architecture in lookup table

According to the proposed architecture in the word lookup table the first 19 bits is used to determine the address of the word in the word lookup table and the next couple of 6-bit combination is for the particular word to be stored in details. Here it is taken 7 characters as an average. Finally the last portion is a 6-bit 0 value, which is usually a combination 000000. Here 6 bit character is used because the word lookup table will consist of dictionary words so general characters and punctuation symbols are needed only and that is

The 1st International Conference on Information Management and Business (IMB2005)

possible using 6-bit character. The example of an entry in the word lookup table is shown below in Figure 3,

| Index Address | Word Stored in Details | Ending signal |
|---|---|---|
| 19-bit Address | a | 000000 |
| 19-bit Address | a n d | 000000 |
| 19-bit Address | P a p e r | 000000 |
| 19-bit Address | R e d u c e | 000000 |

**Figure 3:** Words in lookup table

**2.1.2.1. Special word handling.** The word lookup table has some special situation handling address in the lookup table for many reasons. In text data there may be any name or a constant word or a spelling mistake. As the proposed reduction is a lossless reduction this word will needed to be represented too. But those words will not be found in the word lookup table. In that case a termination signal which is certain valued 19-bit address will be placed in the file. This value will tell the reduction machine that from now until a new word from lookup table entry is not encountered all the data will be considered as an ASCII character which will be 6-bit long. Then after the ASCII values if any lookup table word is encountered it will add a zero valued 6-bit combination which will represent the termination of ASCII values and restart of address values. The address pattern of the special situation handling address is shown in the following table, Table 1.

**Table 1:** Special situation handling addresses

| Index address | Word stored in details | Ending signal |
|---|---|---|
| 01---10 | Termination of address | 000000 |
| 01---01 | Single upper case | 000000 |
| 01---10 | Multiple upper case | 000000 |
| 01---11 | Multiple upper case termination | 000000 |
| 01---00 | Title case | 000000 |
| 01---01 | Single tOGGLE cASE | 000000 |
| 01---10 | Multiple tOGGLE cASE | 000000 |
| 01---10 | Multiple tOGGLE cASE termination | 000000 |

Different case words is also needed to represent in as the same as in the text segment. That is for different type of cases there should be different representation.

By default there will lower case used in general and after punctuation a sentence case will be default. But for other cases "Upper case", "Title Case", "Toggle Case" a lookup table address will be added before that word according to the amount of word of that case. But if any word whose case pattern is not supported by any case symbol then it will be considered as a special constant word and will be treated in that manner.

**2.1.2.2. Punctuation symbol handling.** One of the main exceptions in the word lookup table is that, in the word lookup table for each punctuation sign there is two different entries. One entry consists of only the punctuation sign and the other consists of the punctuations and with a no space protection. Because though after each punctuation a space is generally placed but in some case if after punctuation sign no space is provided in the source file then the machine will count the words before and after punctuation as a single word which was a spelling mistake. But for protecting memory consumption the machine have will handle this problem in the following manner. This format is shown below in Table 2.

**Table 2:** Entry of different punctuation signs

| Index address | Word stored in details | Ending signal |
|---|---|---|
| … | … | 000000 |
| 01---00 | . | 000000 |
| 01---01 | .+ no space | 000000 |
| … | … | 000000 |
| 11---10 | , | 000000 |
| 11---11 | ,+ no space | 000000 |
| … | … | 000000 |

**2.1.3. Memory allocation method.** The text segment will be indexed as a form of 19 bit long addresses consecutively. The white space between each respective word is excluded here. The reason is that, in the binary text stream each 19 bit represents a word and after each word a white space will be automatically added. A question may arise, how multiple space will be solved. The answer is for each character in the ASCII table the lookup table will have three entries. They are single, double and triple alphabetic. In this way the presence of multiple alphabets will be resolved. Here are some examples shown in Figure 3,4,5,6 where LUT stands for lookup table:-

| General | He | is | a | very | good | boy | . | 176 bits |
|---|---|---|---|---|---|---|---|---|
| LUT | 19 | 19 | 19 | 19 | 19 | 19 | 19 | 133 bits |

**Figure 4:** Example 1

| General | Sometimes | I | need | some | help | too | . | 248 bits |
|---|---|---|---|---|---|---|---|---|
| LUT | 19 | 19 | 19 | 19 | 19 | 19 | 19 | 133 bits |

**Figure 5:** Example 2

| General | Although | computers | may | have | basic | similarities | , | 376 bits |
|---|---|---|---|---|---|---|---|---|
| LUT | 19 | 19 | 19 | 19 | 19 | 19 | 19 | 133 bits |

**Figure 6:** Example 3

| General | Several | systematic | tabular | methods | for | machine | reduction | exists | . | 512 bits |
|---|---|---|---|---|---|---|---|---|---|---|
| LUT | 19 | 19 | 19 | 19 | 19 | 19 | 19 | 19 | 19 | 171 bits |

**Figure 7:** Example 4

**2.1.4. Memory space requirement.** Now, to build up a word lookup table of 19-bit and with 75 bit (average) the proposed approach needs a memory space of about,
Space = 2^19 * 67 bits
= 524288 * 67 bits
= 35127296 bits
= 4390912 Bytes
= 4288 Kilo Bytes
= 4.1875 Mega Bytes

Here in 19-bit word lookup table needs only 4.1875 MB memory space to generate and store.

**2.1.5. Lookup table memory allocation.** The word lookup table will be a continuous bit stream like the following pattern. Now as the word lookup table will be a very huge database. Now since the proposed methodology must be faster methodology, to do that the searching is needed to be faster. Thus binary search was a better option. Since, the lookup table is a continuous database so binary search is not possible. But using a Hash table and then doing linear search is a far better solution.

**Table 3:** The Hash Table

| Index address | Starting Character |
|---|---|
| 01---00 | a |
| 01---01 | and |
| … | ..... |
| 10---10 | buy |
| ... | ..... |
| 11---10 | yolk |
| 11---11 | zoo |

The new entries that will be added in the word lookup table will added in the empty entries of the starting character blocks. The word lookup table pattern is shown in Figure 3.8:-

| 11---01 | and | 000000 | 11---10 | andrion | 000000 | 01---11 | daylight | 000000 |
|---|---|---|---|---|---|---|---|---|
| 00---01 | sun | 000000 | 00---10 | sun-bath | 000000 | 01---11 | train | 000000 |

**Figure 8:** How the word lookup table will be stored.

**2.1.6. Data manipulation algorithm.** For the compression of the text data a fixed algorithm will be carried out over the text. General algorithm for converting unreduced text to reduced text is shown in the below:

*Algorithm UnRedToRed( ){*
1. *Read file*
2. *Read character to form a word until empty.*
3. *Finds its appropriate address from Hash table.*
4. *Find the word in Lookup Table.*
5. *If found then*
6. *{*
7. *Check case*
8. *If case = lower then*
9. *Fetch addresses*
10. *else*
11. *{*
12. *Do the case management*
13. *Fetch Address*
14. *}*
15. *Print the address*
16. *}*
17. *else*
18. *{*
19. *Give termination symbol.*

470

   20.    *Start ASCII storage (word)*
   21.    *}*
   22.    *Go to step 1.*
   23.    *End.*
  *}*

General algorithm for converting compressed text to uncompressed text is shown in the below:

*Algorithm RedToUnRed( ) {*
1. *Read file*
2. *Fetch address.*
3. *Check Address status.*
4. *If word then,*
5.   *Print the word.*
6. *If situation handing operator then,*
7.   *Do according to it.*
8.   *Go to step 2.*
9. *End.*
*}*

### 2.2. Text compression

This actual traditional compression is done in this level using the Deflate compression algorithm. This algorithm compresses the text data using both Huffman coding and LZ-77 algorithm. The reduced text file is regenerated as a binary file. The Deflate compression will be carried over the binary file.

### 3. Experimental result

From the Example 1 it is seen that the general size of the text is 176 bits where is by using 19-bit word lookup table the size becomes 133 bits and the percentage of reduction is 24.43%. For Example 2, the result is in general condition 248 bits, but in 19-bit Word Lookup table is 133 bits and the percentage of reduction is 46.37%. Then the result for Example 3, is in general is 376 bits, in 19-bit word lookup table is 133 bits and percentage of reduction is 64.62%. Consequently, according to Example 4 the result is in general is 512 bits, in 19-bit word lookup table is 171 bits and percentage of reduction is 66.60%. Here is another example of the experimental data; the following text segment Example 5 was copied 24 times to make a large text segment. The experimental data Example 5 is shown below:-

"Although computers may have basic similarities, performance will differ markedly between them, and just the same as it does with cars. The PC contains several processes running at the same time, often at different speeds, so a fair amount of coordination is required to ensure that they don't work against each other. Most performance problems arise from bottlenecks between components that are not necessarily the best for a particular job, but a result of compromise between price and performance. Usually, price wins out and you have to work around the problems this creates. The trick to getting the most out of any machine is to make sure that each component is giving of its best, and then eliminate potential bottlenecks between them. You can get a bottleneck simply by having an old piece of equipment that is not designed to work at modern high speed - a computer is only as fast as its slowest component, but bottlenecks can also be due to badly written software."

Another experimental data was created, copying the following text segment in Example 6, 32 times, to make a large text segment. The experimental data Example 6 is shown below:-

"In the current world we have high powerful processors and high capability storages devices not only in the micro computer but also in PDA's. That's why it not difficult to store or to manipulate a file. But it is still difficult to transfer file or data through communication medium. The reason is that the signal capacities of the carriers are not sufficient enough. And this problem is deeply felt in internet communication. In the case of text transfer if we can minimize the text size it will increase the faster portability of the text files. This can be done by indexing the text and by generating a lookup table which will be used to index the text and that will decrease the number of Bytes needed to define a particular text."

The reduction result for the Examples 5 and 6 is shown below in Table 4,

**Table 4:** Size reduction result

| In General Situation | | |
|---|---|---|
| | Example 5 | Example 6 |
| Words | 3984 | 4544 |
| Characters | 19392 | 19328 |
| Characters with white space | 23378 | 23519 |
| Text Size | Bytes 23378 | Bytes 23519 |
| In 21-bit Word Lookup-Table | | |
| Words | 3984 | 4320 |
| Punctuation | 361 | 224 |
| Words with punctuation | 4345 | 4544 |
| Per word text size | bits 19 | bits 19 |
| Text size | Bytes 10320 | Bytes 10792 |
| Text Size Reduction Status | | |
| General situation | 23378 | 23519 |
| 21-bit word lookup-table | 10320 | 10792 |
| Size reduced | 13058 | 12727 |
| Reducing percentage | 55.86% | 54.11% |

We also had experimented the text size for two stories of Leo Tolstoy; and some couple of articles published in local English newspapers. The results for the stories of Tolstoy were 55.91% and 47.32%. For articles in daily newspaper, the results were 40.24%, 55.64%, 64.36%, 52.75%, 49.16% and 56.97%. That is, finally we got an average of 53.4188% reduction rate.

### 3.1. Comparison with other compression method

In general the all compression methods have there compression rate from 12% to highest 50%. But in our method we have found 53% reduction in the starting of the approach without any compression. That is further improvement of the approach will increase the reduction rate.

### 3.2. Comparison with other zip software

In comparison with currently available zip software we found that the following outputs in the case of a same file. The comparison is shown in the Table 5.

**Table 5:** Compression between zip software

| Compression Type | Size |
|---|---|
| Normal | 78.53 KB |
| Proposed Method | 14.38 KB |
| Gzip | 29.61 KB |
| Winzip | 31.27 KB |

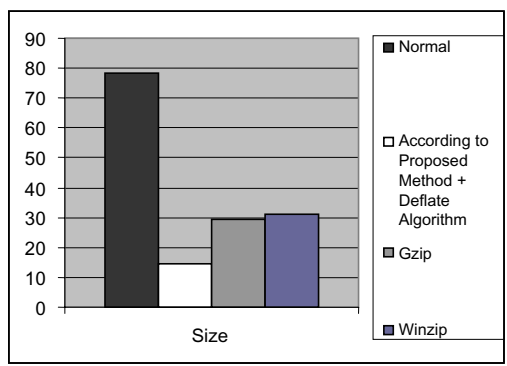

**Chart 1:** Compression between zip software

### 4. Conclusion

In this paper it was desired to provide a whole new compression method. As the world is moving towards the goal of providing highest service at a lowest expense, this method of word lookup table will make any text segment able to use lesser memory space but will not decrease its features rather will increase its usability and portability. It will decrease the memory area occupied by text segment in any type of file, which will make a huge amount of memory area free. And also decrease its transfer time through FTP or SMTP too.

### 5. Limitations and future work

This method may be applied more efficiently if suitable algorithms are applied for determining the address value and doing memory management. Our intention is to use the Deflate algorithm to decrease the index address memory requirement as well the constant words memory requirements.